\documentclass[twocolumn,showpacs,preprintnumbers,amsmath,amssymb]{revtex4}
\usepackage{graphicx}
\usepackage{dcolumn}
\usepackage{bm}
\def\pmb#1{\setbox0=\hbox{$#1$}%
  \kern-.025em\copy0\kern-\wd0
  \kern.05em\copy0\kern-\wd0
  \kern-.025em\raise.0433em\box0}
\def\parb{\pmb{\partial}}
\def\alt{\mathrel{\hbox{\rlap{\hbox{\lower4pt\hbox{$\sim$}}}\hbox{$<$}}}}
\begin{document}

\title{Numerical simulations of stiff fluid gravitational singularities}

\author{Joshua Curtis}
\affiliation{Department of Physics, University of Guelph, Guelph, Ontario,
Canada N1G 2W1}

\author{David Garfinkle}
\email{garfinkl@oakland.edu}
\affiliation{Department of Physics, Oakland University, Rochester, MI 48309}
\begin{abstract}

Numerical simulations of the approach to the singularity in 
spacetimes with stiff fluid matter
are presented here.  The spacetimes examined 
have no symmetries and can be regarded as representing the 
general behavior of singularities in the presence of such matter.  
It is found that the singularity
is spacelike and that as it is approached, the spacetime dynamics 
becomes local and non-oscillatory. 

\end{abstract}
\maketitle

\section{Introduction} 

A longstanding problem in general relativity has been to find the general
behavior of singularities.  Several results, both 
analytical\cite{alanreview}
and numerical\cite{bevreview}
have been obtained.  Though most of the results are for the case where
the spacetimes have one or more symmetries, recent work has been done on
the general case where 
there are no symmetries.\cite{larsandalan,harmonic,dgprl}  
There is a longstanding conjecture\cite{bkl} due to Belinski, Lifschitz and 
Khalatnikov (BKL) that states that the generic singularity is 
spacelike and local.  
This conjecture has been reformulated and
put more precisely by Uggla {\it et al}.\cite{jw} 
The type of local dynamics conjectured by BKL depends on the type of 
matter.  For vacuum and for many types of matter, the BKL conjecture 
is that the local dynamics is oscillatory, corresponding to the 
dynamics of a Bianchi type IX spacetime.  However, for stiff fluid
({\it i.e.} fluid with pressure equal to energy density) the BKL 
conjecture is that the local dynamics is asymptotically velocity
term dominated corresponding to the dynamics of a Bianchi type I 
spacetime.  The vacuum version of the BKL conjecture has been 
supported by the numerical simulations of \cite{dgprl} which show
local and oscillatory dynamics in vacuum spacetimes with no symmetry.
The stiff fluid version of the BKL conjecture has been supported by
the theorem of Andersson and Rendall\cite{larsandalan} which 
shows the local existence in
a neighborhood of the singularity of solutions of the Einstein equations
with stiff fluid matter with the expected asymptotic behavior and with
enough degrees of freedom to be the generic solutions.  What is not
known is whether generic stiff fluid initial data evolves to a solution of
the Andersson and Rendall class.     

To address this issue, we perform numerical simulations of the approach
to the singularity for stiff fluid matter with no symmetries.  Our methods
are those of \cite{dgprl} using the system of \cite{jw}.  Section II 
presents the equations and numerical methods used.  The results are 
given in section III and conclusions in section IV.   

\section{Equations}

The system evolved here is essentially that of reference\cite{jw} but
specialized to the stiff fluid case and 
with a slightly different choice of gauge.  Here the spacetime
is described in terms of a coordinate system ($t,{x^i}$) and a tetrad 
(${{\bf e}_0},{{\bf e}_\alpha}$)  where both the spatial coordinate 
index $i$ and 
the spatial tetrad index $\alpha $ go from 1 to 3.  It is assumed
that ${\bf e}_0$ is hypersurface orthogonal and that the relation between
tetrad and coordinates is of the form
${{\bf e}_0} = {N^{-1}}{\partial _t}$ and  
${{\bf e}_\alpha} =
{{e_\alpha }^i}{\partial _i}$ 
Here $N$ is the lapse and we have chosen the shift to be zero.
We choose the spatial frame $\{ {{\bf e}_\alpha} \}$ to be
Fermi propagated along the integral curves of ${\bf e}_0$.
The commutators of the tetrad components are decomposed as follows:
\begin{eqnarray}
[{{\bf e}_0},{{\bf e}_\alpha}] &=& {{\dot u}_\alpha}{{\bf e}_0}
-(H {{\delta _\alpha}^\beta}
+{{\sigma _\alpha}^\beta})
{{\bf e}_\beta} 
\label{com0i}
\\
\left [ {{\bf e}_\alpha },{{\bf e}_\beta} \right ]  &=&
(2 {a_{[\alpha}}{{\delta _{\beta ]}}^\gamma}
+ {\epsilon _{\alpha \beta \delta }}{n^{\delta \gamma}}){{\bf e}_\gamma}
\label{comij}
\end{eqnarray}
where $n^{\alpha \beta}$ is symmetric, and $\sigma ^{\alpha \beta}$ is 
symmetric and trace free.  Square brackets denote the antisymmetric part of a 
tensor. 

Define ${u^a} \equiv {{e_0}^a}$ and 
${h_{ab}} \equiv {g_{ab}}+{u_a}{u_b}$, that is $u^a$ is the timelike vector
of the tetrad and $h_{ab}$ is the spatial metric corresponding to the
choice of $u^a$ as the time direction.  
Then the stress-energy tensor can be decomposed as
\begin{equation}
{T_{ab}} = \mu {u_a}{u_b} + 2 {q_{(a}}{u_{b)}} + p {h_{ab}} + {\pi _{ab}} 
\label{generalstress}
\end{equation}
where $q_a$ and $\pi _{ab}$ are orthogonal to $u^a$ and where 
$\pi _{ab}$ is symmetric and trace-free.  Round brackets denote the 
symmetric part of a tensor.  

Scale invariant variables are defined as follows: 
\begin{eqnarray}
\{ {\parb_0},{\parb_\alpha} \} \equiv \{ 
{{\bf e}_0},{{\bf e}_\alpha} \} /H 
\label{si1}
\\ 
\{ {{E_\alpha}^i}, {\Sigma _{\alpha \beta }}, {A^\alpha} , 
{N_{\alpha \beta }} \} \equiv \{ {{e_\alpha}^i} , 
{\sigma _{\alpha \beta }} , {a^\alpha}, {n_{\alpha \beta}} \} /H
\label{si2}
\\
q+1 \equiv - {\parb _0} \ln H 
\label{si3}
\\
{r_\alpha} \equiv - {\parb _\alpha} \ln H
\label{si4}
\\
\left \{ \Omega , P , {Q^\alpha}, {\Pi _{\alpha \beta}} \right \}
\equiv \left \{ \mu , p , {q^\alpha} , {\pi _{\alpha \beta}} \right \}
/(3 {H^2})
\label{si5}
\end{eqnarray}

The matter variables are not all independent, because we assume that
the stress-energy is that of a stiff fluid
\begin{equation}
{T_{ab}} = {\tilde \mu} \bigl ( 2 {{\tilde u}_a}{{\tilde u}_b} + {g_{ab}}   
\bigr )
\label{stiffstress}
\end{equation}
Here $\tilde \mu$ is the rest frame energy density of the fluid and 
${\tilde u}^a$ is the fluid four-velocity, which can be decomposed as
${{\tilde u}^a}= \Gamma ({u^a}+{v^a})$ where $v^a$ is orthogonal to 
$u^a$.  Comparison of equations (\ref{generalstress}) and
(\ref{stiffstress}) yields
\begin{eqnarray}
{Q_\alpha} = {\frac {2 \Omega} {G_+}} {v_\alpha}
\\
{\Pi _{\alpha \beta}} = {\frac {2 \Omega} {G_+}} {v_{< \alpha}}{v_{\beta >}}
\\
P = {\frac \Omega  {G_+}} \bigl ( 1 - {\textstyle {\frac 1 3}} {v^2} \bigr )
\end{eqnarray}
Here ${v^2} = {v^\alpha}{v_\alpha}$ and ${G_+} = 1 + {v^2}$ and
angle brackets denote the symmetric trace-free part of a tensor.  
Thus, all scale 
invariant matter variables can be expressed in terms of $\Omega $
and $v^\alpha$.

Finally choose the lapse to be $N={H^{-1}}$.  The relation between
scale invariant frame derivatives and coordinate derivatives is
${\parb _0} ={\partial _t}$ and 
${\parb _\alpha} = {{E_\alpha }^i} {\partial _i}$.
From the Einstein field equations and the conservation of
stress-energy one obtains the following 
evolution equations:
\begin{eqnarray}
{\partial _t} {{E_\alpha}^i} &=& {{F_\alpha}^\beta}{{E_\beta}^i}
\label{ev1}
\\
{\partial _t} {r_\alpha} &=& {{F_\alpha}^\beta}{r_\beta}+{\parb _\alpha}q
\label{ev2}\\
{\partial _t} {A^\alpha} &=& {{F^\alpha}_\beta}{A^\beta}+ 
{\textstyle \frac 1 2}{\parb  _\beta}{\Sigma ^{\alpha \beta}}
\label{ev3}
\\
\nonumber
{\partial _t} {\Sigma ^{\alpha \beta}} &=& (q-2) {\Sigma ^{\alpha \beta}}
- 2 {{N^{<\alpha}}_\gamma}{N^{\beta > \gamma}} + {{N_\gamma }^\gamma}  
{N^{<\alpha \beta >}} 
\\
\nonumber
&+& {\parb ^{<\alpha}}{r^{\beta >}} 
- {\parb ^{<\alpha}}{A^{\beta >}} 
+ 2{r^{<\alpha }}{A^{\beta >}} 
\\
&+& {\epsilon ^{\gamma \delta < \alpha}}({\parb _\gamma } - 2 {A_\gamma})
{{N^{\beta > }}_\delta}  + 3 {\Pi ^{\alpha \beta}} 
\label{ev4}
\\
{\partial _t}{N^{\alpha \beta}} &=& q {N^{\alpha \beta }} + 
2 {{\Sigma ^{(\alpha }}_\delta}{N^{\beta ) \delta }} - 
{\epsilon ^{\gamma \delta (\alpha }}{\parb _\gamma } 
{{\Sigma ^{\beta )}}_\delta}
\label{ev5}
\\
\nonumber
{\partial _t} \Omega &=& (2 q - 1) \Omega - 3 P - {\parb _\alpha} {Q^\alpha}
+ 2 {Q^\alpha}{A_\alpha} 
\\
&-& {\Pi ^{\alpha \beta}} {\Sigma _{\alpha \beta}}
\label{ev6}
\\
\nonumber
{\partial _t} {v^\alpha} &=& {\frac {G_+}{2 {G_-}\Omega}}
\biggl [ \bigl ( {G_-} {{\delta ^\alpha}_\beta} + 2 {v^\alpha}{v_\beta} 
\bigr ) \bigl ( {\partial _t} {Q^\beta} - 2 [q+1] {Q^\beta} \bigr )
\\
&-& 2 {v^\alpha} \bigl ( {\partial _t} \Omega - 2 [q+1] \Omega \bigr ) \biggr ] 
\label{ev7}
\\
\nonumber
{\partial _t} q &=& \left [ 2 (q-2) + {\textstyle \frac 1 3} 
\left ( 2 {A^\alpha } - {r^\alpha}\right ) {\parb _\alpha}
- {\textstyle \frac 1 3} {\parb ^\alpha}{\parb _\alpha}
\right ] q 
\\
\nonumber
&-& {\textstyle \frac 4 3} {\parb _\alpha}{r^\alpha} + 
{\textstyle \frac 8 3}{A^\alpha}{r_\alpha} + {\textstyle \frac 2 3}
{r_\beta}{\parb _\alpha}{\Sigma ^{\alpha \beta}} 
- 2 {\Sigma ^{\alpha \beta}}{W_{\alpha \beta}} 
\\
\nonumber
&+& {\frac 2 {G_+}} \biggl [ 2 \Omega {\Sigma ^{\alpha \beta}} {v_\alpha}
{v_\beta} - 2 (q-2) \Omega + {\partial _t} \Omega  
\\
&-& {\frac {2 \Omega} {G_+}} {v_\alpha} {\partial _t} {v^\alpha} \biggr ] 
\label{ev8}
\end{eqnarray}
Here we are using units where $c = 8 \pi G = 1$.  Furthermore the 
quantities $G_-$,  
$F_{\alpha \beta }$, $W_{\alpha \beta} $ and ${\partial _t} {Q_\alpha}$ 
are given by
\begin{eqnarray}
{G_-} &\equiv & 1 - {v^2}
\\
{F_{\alpha \beta }} &\equiv & q {\delta _{\alpha \beta}} - {\Sigma _{\alpha \beta}}
\\
\nonumber
{W_{\alpha \beta }} &\equiv & {\textstyle \frac 2 3}{N_{\alpha \gamma}}
{{N_\beta}^\gamma} 
- {\textstyle \frac 1 3} {{N^\gamma }_\gamma}
{N_{\alpha \beta }}
+ {\textstyle \frac 1 3} {\parb _\alpha} 
{A_\beta}
\\
&-& {\textstyle \frac 2 3} {\parb _\alpha} {r_\beta}
- {\textstyle \frac 1 3} 
{{\epsilon ^{\gamma \delta }} _\alpha } \left ( {\parb _\gamma}
- 2 {A_\gamma}\right ) {N_{\beta \delta}}  
\\
\nonumber
{\partial _t} {Q_\alpha} &=& 2 (q-1) {Q_\alpha} - {\Sigma _{\alpha \beta}}
{Q^\beta} - {\parb _\alpha} P - {\parb ^\beta} {\Pi _{\alpha \beta}}
\\
\nonumber
&+& (P - \Omega ) {r_\alpha} + {\Pi _{\alpha \beta}} ( 3 {A^\beta} + 
{r^\beta}) 
\\
&+& {\epsilon _{\alpha \beta \gamma}} {N^{\beta \delta}}
{{\Pi _\delta}^\gamma} 
\label{dtqvec}
\end{eqnarray}

In addition to the evolution equations, the variables satisfy constraint 
equations as follows:
\begin{eqnarray}
\nonumber
0 &=& {{({{\cal C}_{\rm com}})}^{\gamma i}}  \equiv 
{\epsilon ^{\alpha \beta \gamma}} 
\bigl ( {\parb _\alpha } {{E_\beta }^i}
 - [{r_\alpha}+{A_\alpha}]{{E_\beta }^i} \bigr ) 
\\
&-& {N^{\gamma \alpha}}
{{E_\alpha }^i} 
\label{cn1}
\\
\nonumber
0 &=& {{\cal C}_{\rm G}} \equiv 1 + {\textstyle \frac 1 3} 
(2 {\parb _\alpha} - 2 {r_\alpha} - 3 {A_\alpha}){A^\alpha} -
{\textstyle \frac 1 6}{N_{\alpha \beta}}{N^{\alpha \beta}}
\\
&+&{\textstyle \frac 1 {12}} {{({{N^\alpha}_\alpha})}^2} -
{\textstyle \frac 1 6} {\Sigma _{\alpha \beta}}{\Sigma ^{\alpha \beta}} 
- \Omega
\label{cn2}
\\
\nonumber
0 &=& {{({{\cal C}_{\rm C}})}^\alpha} \equiv {\parb _\beta} 
{\Sigma ^{\alpha \beta}}+ 2 {r^\alpha} - {{\Sigma ^\alpha}_\beta}
{r^\beta} - 3 {A_\beta}{\Sigma ^{\alpha \beta}}
\\
&-&{\epsilon ^{\alpha \beta \gamma}}{N_{\beta \delta}}
{{\Sigma _\gamma}^\delta}  + 3 {Q^\alpha} 
\label{cn3}
\\
\nonumber
0 &=& {{\cal C}_q} \equiv q - {\textstyle \frac 1 3} {\Sigma ^{\alpha \beta}}
{\Sigma _{\alpha \beta}}+{\textstyle \frac 1 3} {\parb _\alpha }
{r^\alpha} - {\textstyle \frac 2 3}{A_\alpha}{r^\alpha}
\\
&-& {\textstyle {\frac 1 2}} ( \Omega + 3 P )
\label{cn4}
\\
\nonumber
0 &=& {{({{\cal C}_{\rm J}})}^\alpha} \equiv {\parb _\beta} 
{N^{\alpha \beta}} - ({r_\beta}+ 2 {A_\beta}){N^{\alpha \beta}} 
\\
&+& {\epsilon ^{\alpha \beta \gamma}} ({\parb _\beta}{A_\gamma} - 
{r_\beta}{A_\gamma})     
\label{cn5}
\\
0 &=& {{({{\cal C}_{\rm W}})}^\alpha} = {\epsilon ^{\alpha \beta \gamma}}
\bigl ( {\parb _\beta}{r_\gamma} - {A_\beta}{r_\gamma} \bigr ) 
- {N^{\alpha \beta}}{r_\beta}
\label{cn6} 
\end{eqnarray} 

We want a class of initial data satisfying these constraints that 
is general enough for our purposes
but simple enough to find numerically.  
Recall that on an initial data surface, the spatial metric 
$h_{ij}$ and extrinsic curvature $K^{ij}$ must satisfy the constraint
equations
\begin{eqnarray}
{D_i} ({K^{ij}} - K {h^{ij}}) = {q^j}
\label{cmom}
\\
R + {K^2} - {K^{ij}}{K_{ij}} = 2 \mu
\label{cham}
\end{eqnarray}
Here $D_i$ and $R$ are respectively the derivative operator and 
scalar curvature associated with $h_{ij}$ and $\mu$ and $q_i$ are 
the components of the stress-energy tensor given in equation
(\ref{generalstress}).  
We use the York method\cite{jimmy} 
which begins by defining the quantities ${\bar h}_{ij}$ and 
${\bar A}^{ij}$ by
\begin{eqnarray}
{h_{ij}}={\psi ^4} {{\bar h}_{ij}}
\\
{K^{ij}} - {\textstyle {\frac 1 3}} K {h^{ij}} = {\psi ^{-10}} {{\bar A}^{ij}}
\end{eqnarray}
We choose $K$ to be constant, ${\bar h}_{ij}$ to be 
the flat metric $\delta _{ij}$
and $q^i$ to vanish.  With these choices, equations (\ref{cmom}) and
(\ref{cham}) become
\begin{eqnarray}
{\partial _i}{{\bar A}^{ij}} &=& 0
\label{ymom}
\\
{\partial _i}{\partial ^i} \psi &=& \bigl ( {\textstyle {\frac 1 {12}}} 
{K^2} - {\textstyle {\frac 1 4}} \mu \bigr ) {\psi ^5} - 
{\textstyle {\frac 1 8}} {{\bar A}^{ij}}{{\bar A}_{ij}} {\psi ^{-7}} 
\label{yham}
\end{eqnarray}
Here $\partial _i$ is the ordinary derivative with respect to Cartesian
coordinates and indicies are raised and lowered with $\delta _{ij}$.

We choose space to have topology $T^3$ with the Cartesian coordinates
$x, \, y$ and $z$ each going from $0$ to $2 \pi$.  We choose the 
following solution of equation (\ref{ymom})
\begin{eqnarray}
\nonumber
{{\bar A}^{11}}&=&{a_2}\cos y + {a_3} \cos z +{b_2}+{b_3}
\\
\nonumber
{{\bar A}^{22}}&=&{a_1} \cos x - {a_3} \cos z +{b_1}-{b_3}
\\
{{\bar A}^{33}} &=& -{a_1} \cos x - {a_2} \cos y - {b_1} - {b_2}
\end{eqnarray} 
with the off-diagonal components of ${\bar A}^{ij}$ vanishing.  
Here the $a_i$ and $b_i$ are constants.  Note that due to the 
periodicity of the coordinates and the linearity of equation 
(\ref{ymom}) the general solution of equation (\ref{ymom}) is 
a Fourier series.  The solution that we choose is then essentially  
the simplest solution of equation (\ref{ymom}) without symmetries.
The quantity $\mu$ can be freely specified and we choose it to be
\begin{equation}
\mu = {c_0} + {c_1} \cos x + {c_2} \cos y + {c_3} \cos z
\end{equation}
where the $c_i$ are constants.  With these choices for 
${\bar A}^{ij}$ and $\mu$, equation (\ref{yham}) is solved 
numerically (in a manner to be described later) to yield $\psi$
and therefore $h_{ij}$ and $K^{ij}$.  

From this initial data, we must then produce the initial values of
the scale invariant variables.  From equation (\ref{com0i}) it
follows that $H=-K/3$ and since $H$ is constant it then follows
that $r_\alpha$ vanishes.  Since the initial spatial metric is conformally
flat, we choose the initial spatial tetrad vectors by
\begin{equation}
{{E_\alpha}^i}={H^{-1}}{\psi ^{-2}} {{\delta _\alpha}^i}
\end{equation}
It then follows from equation (\ref{comij}) that $N_{\alpha \beta}$
vanishes and that 
\begin{equation}
{A_\alpha} = - 2 {\psi ^{-1}} {\parb _\alpha} \psi
\end{equation}
From equation (\ref{com0i}) it then follows that 
\begin{equation}
{\Sigma ^{\alpha \beta }} = - {H^{-1}}{\psi ^{-6}} {{\delta ^\alpha}_i}
{{\delta ^\beta}_j}{{\bar A}^{ij}}
\end{equation}
while $\Omega$ is given by equation (\ref{si5}) and $q$ by the vanishing of
equation (\ref{cn4}).

The numerical method used is as follows: each spatial direction
corresponds to $n+2$ grid points with spacing $dx=2\pi /n$.  The
variables on grid points $2$ to $n+1$ are evolved using the evolution
equations, while at points $1$ and $n+2$ periodic boundary conditions
are imposed.  The initial data is determined once equation 
(\ref{yham}) is solved.  This is done iteratively as follows:\cite{jim}
Define 
\begin{equation}
S(\psi ) \equiv - 2 \psi + \bigl ( {\textstyle {\frac 1 {12}}}{K^2}
- {\textstyle {\frac 1 4}} \mu \bigr ) {\psi ^5} - 
{\textstyle {\frac 1 8}} {{\bar A}^{ij}}{{\bar A}_{ij}}{\psi ^{-7}}
\end{equation}
Then equation (\ref{yham}) takes the form
$ {\partial ^i}{\partial _i} \psi - 2 \psi = S(\psi )$.  We make 
an initial guess $\psi ^0$ for $\psi $ and solve
using the conjugate gradient
method \cite{Saul}  the equation
\begin{equation}
{\partial ^i}{\partial _i} {\psi ^{k+1}} - 2 {\psi ^{k+1}} = 
S({\psi ^k})
\end{equation}
iterating until $\psi ^k$ satisfies equation (\ref{yham}) to within a 
set tolerance.  

The evolution proceeds using equations 
(\ref{ev1}-\ref{ev8}) with the exception that the term 
$(5-2q) {{\cal C}_q}$ is
added to the right hand side of equation (\ref{ev8}) to prevent the
growth of constraint violating modes and the term 
$-0.6 {{({{\cal C}_{\rm C}})}^\alpha}$ is added to the right hand
side of equation (\ref{ev3}) to make the system well 
posed.\cite{meandcarsten}   Spatial derivatives are 
evaluated using centered differences, and the evolution is done using
a three step iterated Crank-Nicholson method\cite{ICN} (a type of 
predictor-corrector method).  
In equation (\ref{ev8}) the highest spatial derivative term 
is $-{\frac 1 3} {\parb ^\alpha}{\parb _\alpha}q$ which
gives this equation the form 
of a diffusion equation.  Note that diffusion equations can only be
evolved in one direction in time, in this case the negative
direction which corresponds to the approach to the singularity.
Stability of numerical evolution of diffusion equations generally
requires a time step proportional to the square of the spatial step.  
However, the constant of proportionality depends on the coefficient
of the second spatial derivative.  To ensure stability, we define
${E_{\rm max}}$ to be the maximum value of $|{{E_\alpha}^i}|$ (over all
space and over all $\alpha $ and $i$) and then define 
$d{t_1} \equiv - {\frac 1 4} {{(dx/{E_{\rm max}})}^2}$ and 
$d{t_2} \equiv - {\frac 1 8} dx$.  The time step $dt$ is then chosen
to be whichever of $d{t_1}$ and $d{t_2}$ has the smaller magnitude. 

Before presenting numerical results, it is helpful to consider what
behavior to expect as the singularity is approached (that is as
$t \to -\infty$).  First denote the eigenvalues of 
${\Sigma ^\alpha}_\beta$ by (${\Sigma _1},{\Sigma _2},{\Sigma _3}$).
Then suppose that at sufficiently early times the time averages of 
$q - {\Sigma _i}$ are all positive.  Then the time averages of the
eigenvalues of ${F^\alpha}_\beta$ are all positive.  Since we are evolving
in the negative time direction, this should lead (through equation 
(\ref{ev1})) to an exponential decrease in ${E_\alpha}^i$.  However, since
all spatial derivatives appear in the equations in the form
${\parb _\alpha} = {{E_\alpha}^i}{\partial _i}$  we would expect the
spatial derivatives to become negligible.  That is, the approach to 
the singularity is local.  Furthermore, this positivity of the time
averages of the eigenvalues of ${F^\alpha}_\beta$ should lead
(through equations (\ref{ev2}-\ref{ev3})) to exponential decrease
in $r_\alpha$ and $A^\alpha$.  A similar argument applied to equations
(\ref{ev6}) and (\ref{dtqvec})  and using equations (\ref{cn2})
and (\ref{cn4}) 
indicates that as the singularity is approached $Q_\alpha$ should become
negligible, but $\Omega$ should not, and therefore that $v_\alpha$ should
become negligible.  Thus, as the singularity is approached, the system
should be well described by a simplified set of evolution and constraint
equations where spatial derivatives as well as ${r_\alpha}, \, {A^\alpha}$
and $v_\alpha$ are negligible.  Note that the fact that spatial derivatives 
are becoming negligible does $\it not$ mean that the spacetime is becoming 
homogeneous.  Rather the considerable spatial variation is becoming 
a negligible part of the equations of motion since all spatial derivatives
appear multiplied by ${E_\alpha}^i$ which is becoming negligible.  
We now write down this simplified system of evolution and constraint
equations where all these terms are neglected.  
In this
limit equation (\ref{cn3}) implies that the matricies 
${\Sigma ^\alpha}_\beta$ and ${N^\alpha}_\beta$ commute and therefore    
have the same eigenvalues.  The evolution equations for 
$\Sigma ^{\alpha \beta}$ and $N^{\alpha \beta}$ can then be written
as evolution equations for their eigenvalues.  The non-trivial evolution
and constraint equations then become in this limit
\begin{eqnarray}
{\partial _t} {\Sigma _i} &=& (q-2) {\Sigma _i} - 2 {N_i ^2}
+ \left ( {\sum _{k=1} ^3} {N_k} \right ) {N_i} + 
{\textstyle {\frac 2 3}} Y  
\label{trev1}
\\
{\partial _t} {N_i} &=& (q + 2 {\Sigma _i}) {N_i} 
\label{trev2}
\\
{\partial _t} \Omega &=& 2 (q-2) \Omega 
\label{trev3}
\end{eqnarray}
\begin{eqnarray}
0 &=& Y + {\sum _{k=1} ^3} {\Sigma _k ^2} + 6 \Omega - 6
\label{trcn1}
\\
0 &=&  {\sum _{k=1} ^3}{\Sigma_k ^2} + 6 \Omega - 3 q 
\label{trcn2}
\end{eqnarray}
Here $\Sigma _i$ and $N_i$ are the eigenvalues of
${\Sigma ^\alpha}_\beta$ and ${N^\alpha}_\beta$ respectively and 
$Y$ is given by
\begin{equation}
Y = {\sum _{k=1} ^3} {N_k ^2} - {\textstyle {\frac 1 2}}
{{\left ( {\sum _{k=1} ^3} {N_k} \right ) }^2} 
\label{ydef}
\end{equation}
and indicies are not summed over unless explicitly indicated. 

Suppose that the dynamics is in a period (called a Kasner epoch) 
when all the $N_i$ are
negligibly small.  Then it follows from equations (\ref{trcn1})
and (\ref{trcn2}) that $q=2$ and therefore, from equations
(\ref{trev1}) and (\ref{trev3}) that $\Omega$ and
the $\Sigma _i$ are constant.  From equations (\ref{trcn1}) 
and (\ref{trev2}) it follows that there are two possibilities for
a Kasner epoch: (i) all the $\Sigma _i$ are $\ge -1$ in which case
the $N_i$ remain negligible and the Kasner epoch lasts all the way to 
the singularity. (ii) one of the $\Sigma _i$ is $<-1$ in which case
the corresponding $N_i$ grows until it is large enough to bring the
Kasner epoch to an end.  We now look in more detail at possibility (ii).
Let $\Sigma _1$ be the $\Sigma _i$ that is $< -1$.  Then $N_1$ is the
$N_i$ that is growing. 
We are therefore led to examine equations
(\ref{trev1}-\ref{trcn2}) neglecting $N_2$ and $N_3$ but not $N_1$.
In this regime equations (\ref{trev1}) and 
(\ref{trev2}) become 
\begin{eqnarray}
{\partial _t} {\Sigma _1}  = - {S^2} ({\Sigma _1} + 4)
\label{sig1ke}
\\
{\partial _t} {\Sigma _2}  = - {S^2} ({\Sigma _2} - 2)
\label{sig2ke}
\\
{\partial _t} {\Sigma _3}  = - {S^2} ({\Sigma _3} - 2)
\label{sig3ke}
\\
{\partial _t} \Omega  = - 2 {S^2} \Omega 
\label{omegake}
\end{eqnarray}
where $S \equiv {N_1}/{\sqrt 6}$.  
Define $Z \equiv 4 + {\Sigma _1}$.  Then equation (\ref{sig1ke})
becomes $ {\partial _t} Z = - {S^2} Z$ while from equations 
(\ref{sig2ke}) and (\ref{sig3ke}) it follows that there are 
constants $c_2$ and $c_3$ with ${c_2}+{c_3}=-1$ such that
${\Sigma _2} = 2 + {c_2} Z$ and ${\Sigma _3}=2 + {c_3} Z $.  Similarly,
it follows from equation (\ref{omegake}) that there is a constant
$c_4$ such that $\Omega ={c_4} {Z^2}$.  Finally, it then follows from
equation (\ref{trcn1}) that $Z$ satisfies the equation of motion
\begin{equation}
{\partial _t} Z = \left [ {\textstyle {\frac 1 6}} (4 - {\alpha ^2})
{Z^2} - 4 Z + 6 \right ] Z
\end{equation}   
where ${\alpha ^2} = 1 - [12 {c_4} + {{({c_2}-{c_3})}^2}]$.  Note 
that the quantity in square brackets vanishes at $Z_+$ and $Z_-$
where $ {Z_\pm}=6/(2 \mp \alpha )$.  Therefore the dynamics is a 
``bounce'' from a Kasner epoch corresponding to $Z_-$ to one 
corresponding to $Z_+$.  Use a minus subscript to denote a quantity
before the bounce and a plus subscript to denote a quantity after 
the bounce.  We then have ${\Omega _+}/{\Omega _-}={{({Z_+}/{Z_-})}^2}$.
However, from the definition of $Z$ it follows that 
${Z_-}=4+{\Sigma_{1-}}$ while from the definition of $Z_\pm$ it follows
that $(1/{Z_+})+(1/{Z_-})=2/3$.   We then find the following ``bounce rule''
relating a quantity after the bounce to quantities before the bounce.  
\begin{equation}
{\Omega _+} = {\Omega _-} {{\left ( {\frac 3  {5 + 2 {\Sigma _{1-}}}}
\right ) }^2}
\label{rlbounce}
\end{equation}    
Note that from the bounce rule it follows that $\Omega$ increases at each 
bounce.  Furthermore, it follows from equation (\ref{trcn1}) 
(and from the fact that  $\Sigma _{\alpha \beta}$  is trace-free) that
the minimum possible value for a $\Sigma _i$ during a Kasner epoch is
$-2 {\sqrt {1 - \Omega}}$ and therefore that no further bounces can 
happen once $\Omega > 3/4$ (though bounces may cease at lower values of 
$\Omega$).  Thus we expect that at each spatial point there is a last
bounce followed by a Kasner epoch that lasts all the way to the singularity.
In other words, we expect the approach to the singularity to be of the 
Andersson and Rendall class.

\section{results}

All runs were done in double precision on a SunBlade 2000 with $n=50$ 
(except for a convergence test which also used $n=25$).  The 
equations were evolved from $t=0$ to $t=-90$.  For the initial data,
the trace of the extrinsic curvature was $-1$ corresponding to an
initial value of $1/3$ for $H$.  The constants ${a_i}, \, {b_i}$
and $c_i$ characterizing the initial data were
\begin{eqnarray}
{a_i}=(0.2,0.1,0.04)
\\
{b_i}=(1.7,0.1,0)
\\
{c_i} = (0.005,0.005,0.005)
\end{eqnarray}
and the constant $c_0$ was $0.02$.  

We would like to know whether ${{E_\alpha}^i}, \, {r_\alpha}, \, {A^\alpha}$
and $v^\alpha$ become negligible as the singularity is approached.  In 
figure 1 are plotted the maximum values (over all space, $\alpha$ and $i$)
of $\ln | {{E_\alpha }^i}|, \, \ln |{r_\alpha}|, \, \ln |{A^\alpha}|$
and $\ln |{v^\alpha}|$.  Note that after a certain amount of evolution, all 
these quantities steadily decrease.  This indicates that after a certain
amount of time spatial derivatives become negligible in the eqautions of motion
and that the approximation considered at the end of the previous section
becomes valid.  
\begin{figure}
\includegraphics[scale=0.6]{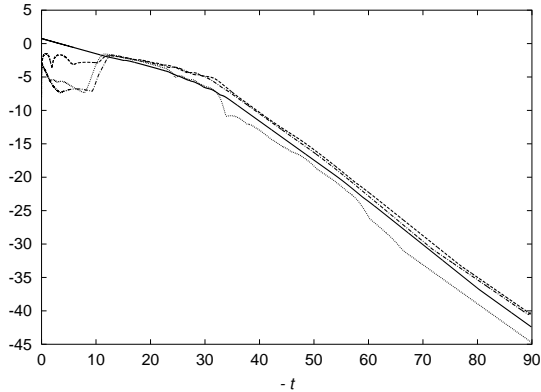}
\caption{\label{fig1} maximum values of 
$\ln |{{E_\alpha }^i}|$ (solid line), $\ln |{r_\alpha}|$ (dotted line)  
$\ln |{A^\alpha}|$ (dot-dashed line) and $\ln |{v^\alpha}|$  
(dashed line) {\it vs} $-t$}
\end{figure}
\begin{figure}
\includegraphics[scale=0.6]{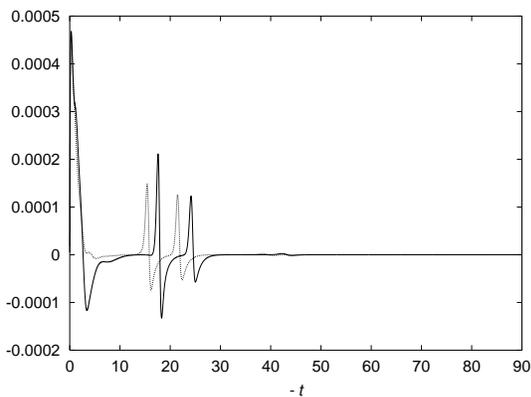}
\caption{\label{fig2} 
$4 {{\cal C}_q}$  {\it vs} $-t$ for the $n=50$ run (solid line)
and ${\cal C}_q$ {\it vs} $-t$ for the $n=25$ run (dotted line)}
\end{figure}
\begin{figure}
\includegraphics[scale=0.6]{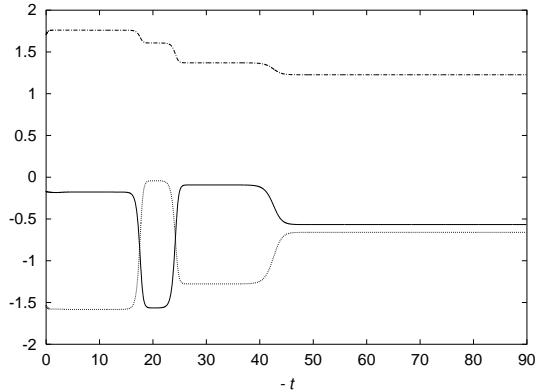}
\caption{\label{fig3} 
components of $\Sigma _{\alpha \beta}$ in the asymptotic frame  {\it vs} $-t$
$\Sigma _1$ (solid line), $\Sigma _2$ (dotted line) and
$\Sigma _3$ (dot-dashed line)}
\end{figure}
\begin{figure}
\includegraphics[scale=0.6]{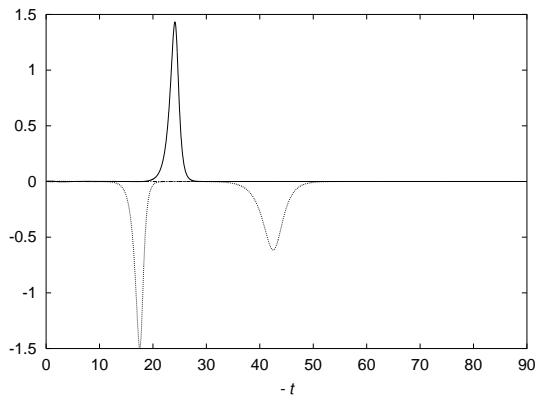}
\caption{\label{fig4} 
components of $N_{\alpha \beta}$  in the asymptotic frame {\it vs} $-t$
$N_1$ (solid line), $N_2$ (dotted line) and $N_3$ (dot-dashed line)}
\end{figure}
\begin{figure}
\includegraphics[scale=0.6]{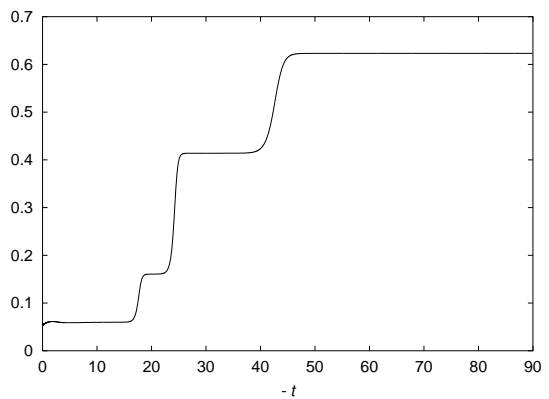}
\caption{\label{fig5} 
$\Omega $  {\it vs} $-t$}
\end{figure}

It then follows that the interesting part of the dynamics is the 
development of the variables at a single point as a function of time.  
We now present data of that form.  The behavior at the spatial point chosen
is typical.

The results of a  convergence test are plotted in figure 2.  
Here what is plotted is 
$4 {{\cal C}_q}$ for the $n=50$ run (solid line) and 
${{\cal C}_q}$ for the $n=25$ run (dotted line).  Both quantities
are plotted {\it vs} $-t$.  Note that the two curves roughly agree in 
magnitude but become out of sync in time.  This
indicates second order convergence but with the system having sensitive
dependence on initial conditions.  Similar results were obtained for the
other constraints.  

Figures 3 and 4 show respectively the diagonal components of 
$\Sigma _{\alpha \beta}$ and $N_{\alpha \beta}$ in the asymptotic frame, 
{\it i.e.} the frame of the eigenvectors that ${\Sigma ^\alpha}_ \beta$ 
has at the end of the evolution.  For that part of the evolution where 
the approximation made at the end of the previous section is valid, these
diagonal components are the eigenvalues of ${\Sigma ^\alpha }_\beta$
and ${N^\alpha}_\beta$ respectively.  Note that the dynamics of the
eigenvalues of ${\Sigma ^\alpha }_\beta$ consists of epochs where they
are apporoximately constant (Kasner epochs) punctuated by short bounces.
Furthermore the components of $N_{\alpha \beta}$ are negligible except
at the bounces.  Also note that there is a last bounce and that this 
coincides with the most negative eigenvalue of ${\Sigma ^\alpha}_\beta$
becoming greater than $-1$.    
   
In figure 5 is plotted $\Omega$ {\it vs} $-t$.  Note that the behavior of 
$\Omega$ is a sequence of constant values that are punctuated by short
bounces and that the bounces in $\Omega$ occur at the same times as the
bounces in $\Sigma _{\alpha \beta}$.  The sequence of the values of
$\Omega$ is 0.05956, 0.1607, 0.4139, 0.6231,  while the corresponding 
values of the most negative
eigenvalues of ${\Sigma ^\alpha} _\beta$ are 
-1.583, -1.565, -1.277, -0.6596.  These values obey
the ``bounce rule'' of equation (\ref{rlbounce}).   

\section{conclusions}

These simulations support the expected picture for the approach to the
generic singularity in a spacetime where the matter is a stiff fluid. 
As the singularity is approached the terms in the equations of motion
involving spatial derivatives become negligible.  The dynamics at 
each spatial point consists of a sequence of Kasner epochs punctuated
by short bounces.  The sequence of values of $\Omega$ obeys the expected
bounce rule.  There is a last bounce, after which the dynamics is described
by a single Kasner epoch all the way to the singularity, thus yielding a
spacetime in the class of reference \cite{larsandalan}.

What remains to be done is to treat the approach to the singularity for
non-stiff fluids.  Here the behavior that is expected is quite different.
The BKL conjecture is that the matter will become negligible and the
dynamics of the gravitational variables as the singularity is approached 
will be that of vacuum
spacetimes.  The formalism of \cite{jw} includes a class of non-stiff
fluids, and the resulting equations are similar to those of the stiff
fluid case.  Nonetheless, the numerical methods of this paper are not
adequate to treat the case of non-stiff fluids.  That is because in
non-stiff fluids shock waves form, while the numerical methods of the
present paper are appropriate for smooth solutions.  A shock capturing 
method would be appropriate to treat the  
non-stiff fluid case.

\section{acknowledgements}

This work was partially supported by a grant from the National Science and 
Engineering Research Council of Canada and by NSF grant
PHY-0456655 to Oakland University.

\end{document}